\documentclass[journal=jpcbfk,manuscript=article]{achemso}

\usepackage{upgreek}
\usepackage{graphicx}
\usepackage{amssymb}
\usepackage{graphicx}
\usepackage{natbib}
\usepackage{amsmath}
\usepackage{amsfonts}
\usepackage{color}
\providecommand{\U}[1]{\protect\rule{.1in}{.1in}}
\DeclareUnicodeCharacter{2212}{-}

\title{Nanocrystallites Modulate Intermolecular Interactions in Cryoprotected Protein Solutions}

\author{Mariia Filianina}
\affiliation{Department of Physics, AlbaNova University Center, Stockholm University, S-106 91 Stockholm, Sweden}
\email{mariia.filianina@fysik.su.se}

\author{Maddalena Bin}
\affiliation{Department of Physics, AlbaNova University Center, Stockholm University, S-106 91 Stockholm, Sweden}

\author{Sharon Berkowicz}
\affiliation{Department of Physics, AlbaNova University Center, Stockholm University, S-106 91 Stockholm, Sweden}

\author{Mario Reiser}
\affiliation{Department of Physics, AlbaNova University Center, Stockholm University, S-106 91 Stockholm, Sweden}

\author{Hailong Li}
\affiliation{Department of Physics, AlbaNova University Center, Stockholm University, S-106 91 Stockholm, Sweden}
\alsoaffiliation{Max Plank Institute for Polymer Research, Ackermannweg 10, 55128 Mainz, Germany}

\author{Sonja Timmermann}
\affiliation{Department of Physics, Universit\"at Siegen, Walter-Flex-Strasse 3, 57072 Siegen, Germany}

\author{Malte Blankenburg}
\affiliation{Deutsches Elektronen-Synchrotron (DESY), Notkestrasse 85, 22607 Hamburg, Germany}

\author{Katrin Amann-Winkel}
\affiliation{Department of Physics, AlbaNova University Center, Stockholm University, S-106 91 Stockholm, Sweden}
\alsoaffiliation{Max Plank Institute for Polymer Research, Ackermannweg 10, 55128 Mainz, Germany}
\alsoaffiliation{Institute of Physics, Johannes Gutenberg University, 55128 Mainz, Germany}

\author{Christian Gutt}
\affiliation{Department of Physics, Universit\"at Siegen, Walter-Flex-Strasse 3, 57072 Siegen, Germany}

\author{Fivos Perakis}
\affiliation{Department of Physics, AlbaNova University Center, Stockholm University, S-106 91 Stockholm, Sweden}
\email{f.perakis@fysik.su.se}

\begin{document}
\date{}

\begin{abstract}

Studying protein interactions at low temperatures has important implications for optimizing cryostorage processes of biological tissue, food, and protein-based drugs. One of the major challenges is related to the formation of ice nanocrystals which can occur even in the presence of cryoprotectants and can lead to protein denaturation. Here, using a combination of small- and wide-angle X-ray scattering (SAXS and WAXS), we investigate the structural evolution of concentrated Lysozyme solutions in a cryoprotected glycerol-water mixture upon cooling from room temperature ($T=300$~K) down to cryogenic temperatures ($T=195$~K). Upon cooling, we observe a transition near the melting temperature of the solution ($T\approx245$~K), which manifests both in the temperature dependence of the scattering intensity peak position reflecting protein-protein length scales (SAXS) and the interatomic distances within the solvent (WAXS). Upon thermal cycling, a hysteresis is observed in the scattering intensity, which is attributed to the formation of nanocrystallites in the order of $10$~nm. The experimental data is well described by the two-Yukawa model, which indicates temperature-dependent changes in the short-range attraction of the protein-protein interaction potential. Our results demonstrate that the nanocrystal growth yields effectively stronger protein-protein attraction and influences the protein pair distribution function beyond the first coordination shell.

\end{abstract}

\newpage

\section{INTRODUCTION}

Organisms that thrive in cold environments have evolved unique strategies to enable survival, including the accumulation of osmolytes and the usage of specialized cryoprotectants~\cite{storey_freeze_1988}. However, protein functionality at low temperatures remains poorly understood due to experimental challenges related to ice formation, which limits the investigation of biomolecules in deeply supercooled environments. Understanding the effect of cryoprotectants in low-temperature protein solutions is important for elucidating the combined effect of the solutes on the freezing point depression and has significant implications for biotechnical cryostorage applications~\cite{hubalek_protectants_2003}.

At low temperatures and below what is known as the protein dynamic transition ($T_d$ $\approx$ 230~K), proteins are believed to lose their conformational flexibility required for biological function~\cite{doster_dynamical_1989,ringe_glass_2003}. Although the origin of this effect is still debated, it has been reported for many biopolymers and is now accepted as a generic feature of hydrated proteins, while it is absent in dehydrated systems~\cite{doster_dynamical_2010, fitter_temperature_1999}. Experimental studies suggest this effect stems from the crossover in proteins' intrinsic dynamics from harmonic to anharmonic motions above $T_d$~\cite{doster_dynamical_1989} and that temperature-induced phenomena in the hydration shell and the bulk solvent play a crucial role~\cite{chen_observation_2006,cordone_harmonic_1999,paciaroni_effect_2002,tsai_molecular_2000, caliskan_dynamic_2002,fenimore_concepts_2013,frauenfelder_unified_2009,fenimore_bulk-solvent_2004}. An important observation is that the transition temperature, observed in the mean square displacement amplitude, depends on the properties and the amount of the molecules surrounding the protein surface~\cite{cordone_harmonic_1999,paciaroni_effect_2002}. 
Such sensitivity of biomolecules to the solvent implies the possibility to control both the onset and the amplitude of the protein anharmonic motions related to the dynamical transition by choosing a suitable environment~\cite{tsai_molecular_2000,cordone_harmonic_1999}, which can have various practical consequences in connection with the development of biological cryogenic techniques~\cite{hubalek_protectants_2003}.

This aspect further emphasises the importance to accurately account for the critical phenomena in the solvent itself. For example, glycerol is widely employed in studies of low-temperature protein dynamics~\cite{caliskan_protein_2003,caliskan_protein_2004,ronsin_preferential_2017,jansson_role_2011,dirama_coupling_2005} and structure~\cite{hirai_direct_2018} due to its ability to induce strong frustration against water crystallization~\cite{tanaka_liquidliquid_2020, murata_liquidliquid_2012}. This effect arises from the observation that the glycerol molecules affect the local structure and hydrogen bonding of water~\cite{jansson_role_2011,farnum_effect_1999} and suppress the tetrahedral component~\cite{daschakraborty_how_2018}. Furthermore, glycerol aqueous solutions have been hypothesised to exhibit a liquid-liquid transition~\cite{murata_liquidliquid_2012,murata_general_2013}, although the physical understanding for this phenomenon is still debated. Early experimental studies ascribe the observed low temperature transition in glycerol-water mixtures to the genuine liquid-liquid transition~\cite{murata_liquidliquid_2012,murata_general_2013}, whereas followup investigations suggest that the observed transition may be due to nanocrystallite formation, which occurs due to the demixing of glycerol-water mixtures at lower temperatures. Importantly, these effects in glycerol can be experimentally observed in a narrow range of the glycerol concentration from 15 mol$\%$ to 28 mol$\%$~\cite{bachler_glass_2016,bachler_glass_2019,hayashi_relaxation_2005,hayashi_ice_2005}, which is close to those typically used in cryopreservation applications for biological molecules, cells, and embryos~\cite{hubalek_protectants_2003,yoshimura_cryopreservation_2019}. Hence, a unified understanding of the mutual effects of the solvent on the proteins and vice versa at low temperature remains of fundamental importance.

Here, we focus on the structural investigation of cryoprotected protein solutions in a wide temperature range.  
Using small- and wide-angle X-ray scattering (SAXS/WAXS), we study Lysozyme in glycerol-water solutions (23 mol$\%$ glycerol). The combination of SAXS/WAXS allows us to simultaneously follow the changes in the intermolecular protein-protein and interactomic interactions within the solvent. Measurements are performed over a broad range of temperatures including thermal cycles cooling from room temperature ($T=300$~K) down to $T=195$~K and warming back up. We model the measured SAXS intensities with the two-Yukawa potential describing the protein-protein interactions to elucidate the origin of the observed temperature-induced transitions.

\section{METHODS}
\subsection*{Materials and sample preparation}
Lysozyme from hen egg white (14.3 kDa) was purchased from Sigma-Aldrich (L6876) and was used without further purification. The protein powder was dissolved in a 23~mol$\%$ (corresponding to 55~vol$\%$ or 60~w\%)
glycerol-water solution with protein concentrations of 10~mg/ml and 200~mg/ml. 
The resulting solutions were filled in quartz capillaries with 1.5~mm diameter for X-ray scattering studies. \\

\subsection*{Experimental X-ray parameters}
Small- and wide-angle X-ray scattering experiments (SAXS and WAXS) were carried out at the High Energy X-Ray Diffraction beamline P21.1 at PETRA III (DESY, Hamburg), using the experimental parameters as detailed in Table~\ref{tab:exp-param}. 
The measured two-dimensional (2D) X-ray scattering patterns were azimuthally averaged to obtain the $I(Q)$ curves. The resulting scattering curves were corrected for solvent and background scattering by subtracting the scattering intensity measured on a capillary filled with the glycerol-water solvent of the same concentration.
A Linkam scientific instruments stage (model HFSX350) was used to control and vary the sample temperature within a broad range from $T=300$~K to $T=195$~K. For all measurements the temperature was varied with the rate of 4~K/min and the SAXS/WAXS scattering patterns were measured simultaneously and continuously as the temperature was varied. Measurements of the temperature cycles were performed on different capillaries filled with identical samples of the same solution. Each temperature cycle was measured on a single spot of the sample. 

\begin{table}[tbhp]
  \centering
  \caption{The experimental X-ray parameters used for the experiment, including the photon energy, beam size, flux, sample environment, detector and sample-detector distance (SDD) for SAXS and WAXS geometries. }
  \label{tab:exp-param}
  \begin{tabular}{lcc}
    \hline
    Photon energy (keV)  & 52.5   \\
    Beam size ($\upmu$m$^2$) & $500\times500$  \\
    Flux (x$10^9$\,ph/s)& 12.5  \\
    Sample environment  & Linkam stage   \\
    SAXS detector  &    Pilatus3~X~CdTe~2M   \\
    SAXS SDD (m) &  14.6    \\
    WAXS detector   &   Varex~XRD4343CT\\
    WAXS SDD (m) & 1.0    \\
    \hline
  \end{tabular}
\end{table}

\subsection*{Data analysis and modeling}

The scattering patterns acquired with the 2D detectors were normalized by the intensity of the transmitted beam, azimuthally averaged using the \textsc{pyFAI} python library~\cite{ashiotis_fast_2015} followed by the subtraction of the solvent scattering.

The scattering intensity $I(Q)$ as a function of the momentum transfer $Q = \frac{4 \pi}{\lambda} \sin{\theta}$, where $2\theta$ is the scattering angle and $\lambda$ is the wavelength of the X-ray source, was modelled by the following expression: 

\begin{equation}
    I(Q) = \phi V \Delta \rho^2 \left< P(Q) \right> S(Q) + c.
    \label{eq:iofq}
\end{equation}

Here, $\rho$ is the averaged contrast term, $V$ and $\phi$ are the volume and the volume fraction of an individual protein, respectively. The orientation-averaged form factor $\left< P(Q) \right>$ is related to the protein size and structure, while the structure factor $S(Q)$ provides information about the protein-protein interactions, which for dilute solutions of non-interacting proteins corresponds to $S(Q) \approx 1$. $c$ is the background offset, which was determined based on the large $Q$ asymptotic value at about $Q \approx 3.5$ nm$^{-1}$. The volume fraction is $\phi \approx 0.15$ for the 200~mg/ml Lysozyme solution and was used as a fixed parameter for the data analysis. 

For Lysozyme in solution, the form factor can be described by that of an ellipsoid of revolution~\cite{shukla_absence_2008, liu_cluster_2005, moller_effect_2012} as
\begin{equation}
P(Q) = \int_{0}^{1} \frac{j_1 \left( Q \sqrt{r_{a}^2 + x^2(r_{b}^2 - r_{a}^2)} \right)^2 }{ \left( Q \sqrt{r_{a}^2 + x^2(r_{b}^2 - r_{a}^2)} \right) ^4}\,\text{d}x \,, 
    \label{eq:formfactor}
\end{equation}
where  $r_{a}$ and $r_{b}$ are the ellipsoid semi axes. In this work, the experimental $I(Q)$ curves for the lowest protein concentration were fitted with a normalized radially averaged scattering function of an ellipsoid with a fixed aspect ratio of $r_{a}/r_{b} = 1.5$~\cite{shukla_absence_2008, liu_lysozyme_2011} calculated with \textsc{jscatter}~\cite{biehl_jscatter_2019}, while keeping the major semi axis as a fitting parameter.

The $S(Q)$ is related to the effective interaction potential $U(r)$ through the direct correlation function, which in turn can be obtained within the mean spherical approximation~\cite{liu_cluster_2005, liu_lysozyme_2011}. 
Here, we use the two-Yukawa (TY) potential to describe the protein-protein interaction, which has been previously used successfully to describe the Lysozyme structure factor at highly concentrated conditions~\cite{shukla_absence_2008,liu_lysozyme_2011,riest_short-time_2018}. The TY potential
comprises a short-range attraction and a long-range repulsion term as follows

\begin{equation}
\frac{U(r)}{k_{B}T} = 
\begin{cases}
\infty \,, & \text{where $r < \sigma $}\\
-K_{1}\frac{e^{-Z_{1}(r/\sigma-1)}}{r/\sigma} + K_{2}\frac{e^{-Z_{2}(r/\sigma-1)}}{r/\sigma} \,, &\text{where $r \geq \sigma$}
\end{cases}
\label{eq:2Y}
\end{equation}

Here, $k_{B}$ is the Boltzmann constant, $T$ is the temperature, $r$ is the protein-protein distance and the effective diameter is $\sigma$. Moreover, $Z_1$ and $Z_2$ determine the range of the attractive and repulsive Yukawa potential terms in units of $\sigma$, respectively, while $K_1$ and $K_2$ correspond to the attractive and repulsive interaction strength in units of $k_{B}T$.

In the modeling of interactions, the attraction strength $K_1$ is used as a fitting parameter, while the parameters $Z_1 = 21 $, $K_2 = 3.2$, and $Z_2 = 3.5$ are fixed~\cite{shukla_absence_2008}. The shape parameters, such as the particle diameter which in terms of the ellipsoid parameters is $\sigma = 2(r_ar_b^2)^{1/3}$, were obtained independently from the form factor analysis.

\section{RESULTS AND DISCUSSION}

\begin{figure}[htbp]
  \centering
  \includegraphics[width=0.7\columnwidth]{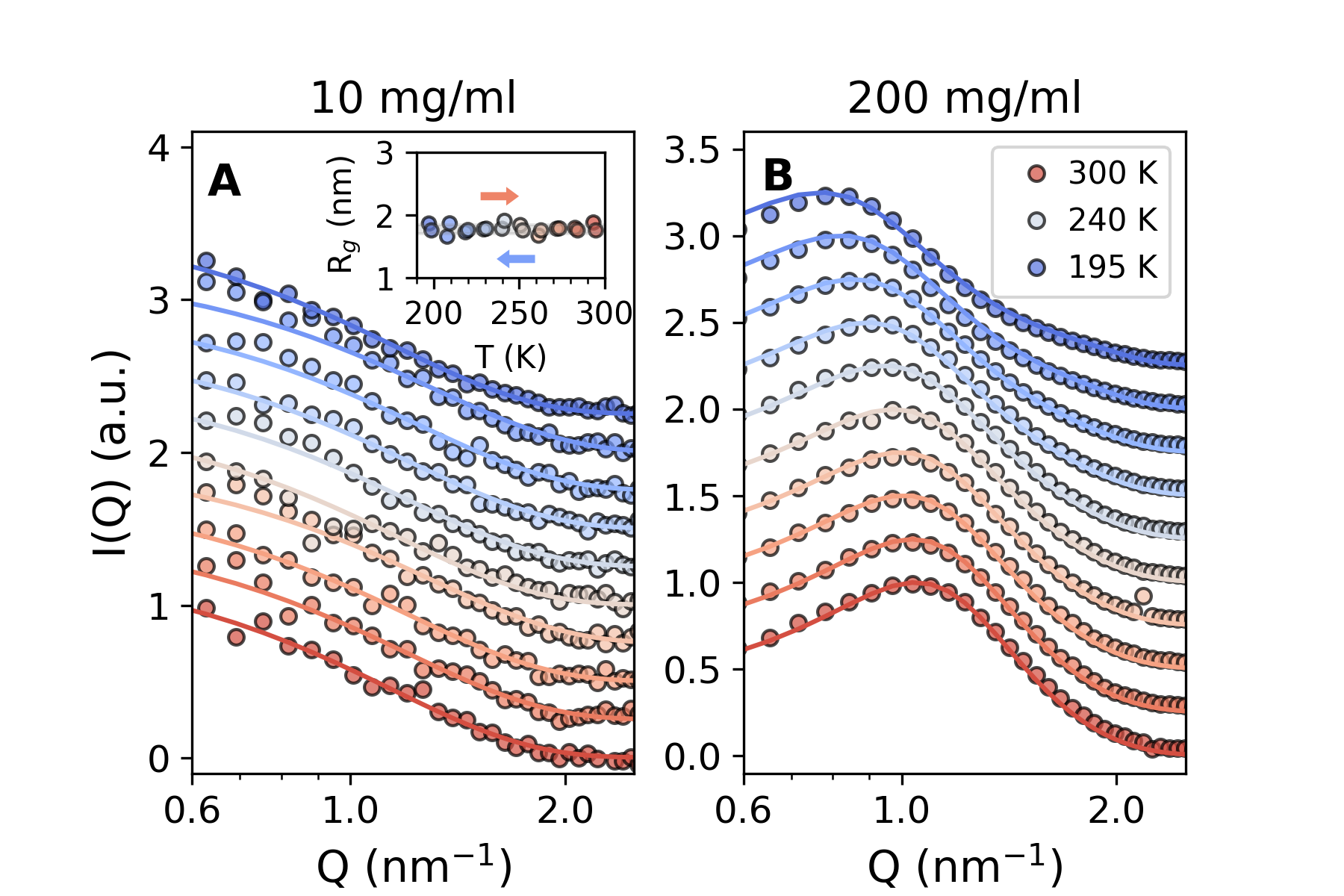}
  \caption{Small-angle X-ray scattering (SAXS) intensities of 10~mg/ml (panel A) and 200~mg/ml (panel B) Lysozyme in 23~mol$\%$ glycerol-water solution as a function of temperature while cooling down from room temperature $T=300$~K (red) to $T=195$~K (blue). Here, an offset has been added to facilitate the comparison. The symbols represent the experimental data while the solid lines are the fits from the model. The inset shows the radius of gyration $R_{g}$ as a function of temperature as extracted from the fits in panel A during cooling and heating.} 
  \label{fig:iqs}
\end{figure}

Figure~\ref{fig:iqs} shows the variation of the SAXS scattering intensity measured for a dilute (panel A, 10~mg/ml) and concentrated (panel B, 200~mg/ml) Lysozyme glycerol-water solution upon cooling the sample from room temperature ($T=300$~K) down to $T=195$~K. The scattering patterns were recorded continuously during the temperature sweep with a cooling rate of 4 K/min and here we show the $I(Q)$ curves averaged over the temperature interval of $\approx$~10~K.    

As seen in Fig.~\ref{fig:iqs}A, the scattering patterns at the lowest protein concentration ($10$~mg/ml) exhibit negligible interference effects and are adequately fitted with the computed scattering intensity of an ellipsoid of revolution with a fixed aspect ratio of $r_{a}/r_{b} = 1.5$. The resulting fits to the data are shown as solid lines in Fig.~\ref{fig:iqs}A for various temperatures. As shown in the inset in Fig.~\ref{fig:iqs}A, no significant variation is observed in the radius of gyration $R_\text{g}$ in the entire temperature range, suggesting that we do not observe any significant changes in the protein size related to cold denaturation~\cite{kim_computational_2016,kozuch_low_2019}. For all temperatures the gyration radii are determined $R_\text{g} = 1.78\pm 0.07$~nm which reasonably agrees with the literature values for Lysozyme~\cite{shukla_absence_2008,cardinaux_modeling_2007}. In the presence of glycerol, which acts as a cryoprotectant stabilizing proteins at low temperatures, Lysozyme is known to slightly compactify~\cite{bonincontro_dielectric_2003}, while still remaining in its quasi-native state\cite{joshi_quasi-native_2020}.  

\begin{figure}[htbp]
  \centering
  \includegraphics[width=0.7\columnwidth]{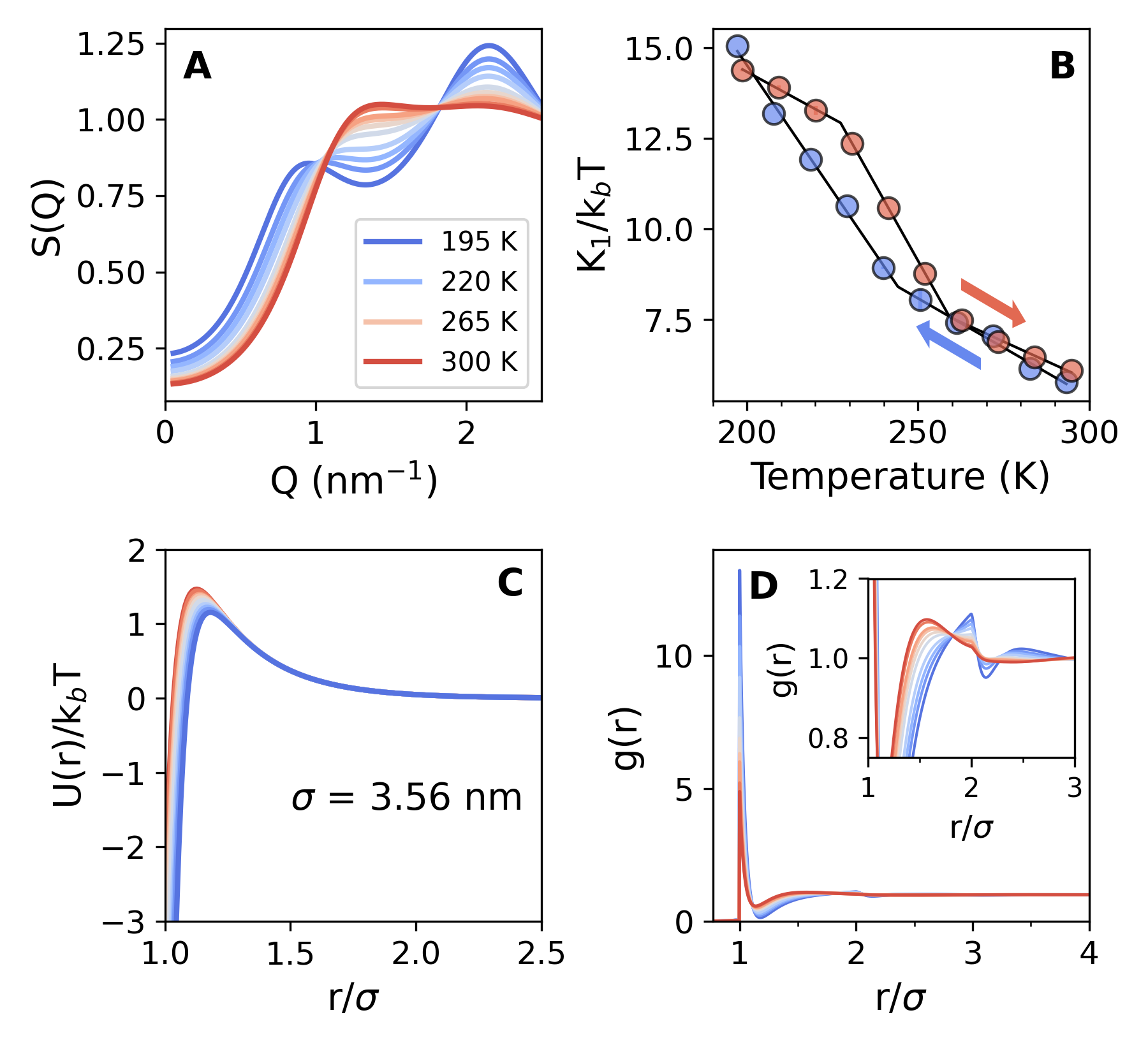}
  \caption{(A) The protein-protein structure factor $S(Q)$ obtained from the fits using the two-Yukawa (TY) model shown in Fig.~\ref{fig:iqs}B at different temperatures upon cooling. (B) Temperature dependence of the attraction strength parameter $K_1$ extracted from fitting the SAXS curves upon cooling down (blue) and warming up (red). The other TY parameters were fixed to $Z_1 = 21$, $Z_2 = 3.5$, and $K_2 = 3.2$ k$_\text{B}$T. (C) Temperature evolution of TY potential upon cooling down. (D) The pair distribution function $g(r)$ derived from the modeled structure factors shown in A. The inset shows a detail of the $g(r)$ second coordination shell.  }
  \label{fig:model}
\end{figure}

The scattering patterns for higher protein concentration (200~mg/ml) shown in Fig.~\ref{fig:iqs}B exhibit an interference peak close to $Q\approx1$ nm$^{-1}$ at room temperature, related to the intermolecular protein structure factor. With decreasing temperature, the peak shifts to lower $Q$ values in the entire temperature range accessed here. This behaviour is consistent with the temperature trends reported for Lysozyme in buffer solutions at ambient conditions in previous studies~\cite{shukla_absence_2008,stradner_equilibrium_2004, moller_effect_2012}. 
The solid lines represent the fits of the $I(Q)$ curves using the TY model in order to extract the structure factor, as described in Methods. The calculated structure factor $S(Q)$ using the best fit parameters is shown in Fig.~\ref{fig:model}A as well as the temperature dependence of the fit parameter in Fig.~\ref{fig:model}B which are discussed in the following paragraphs.

\begin{figure}[htbp]
  \centering
 
  \includegraphics[width=0.7\columnwidth]{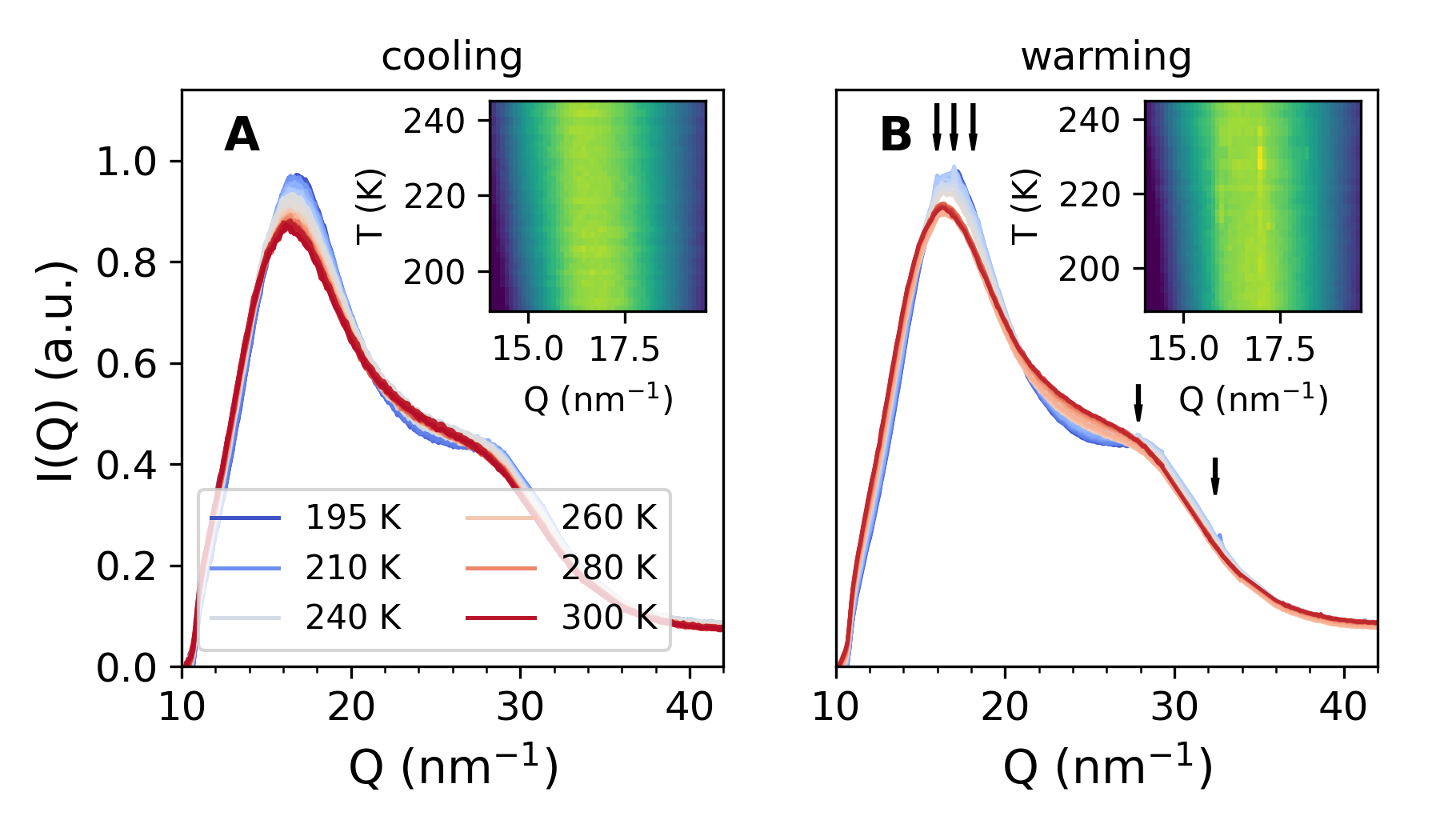}
  \includegraphics[width=0.7\columnwidth]{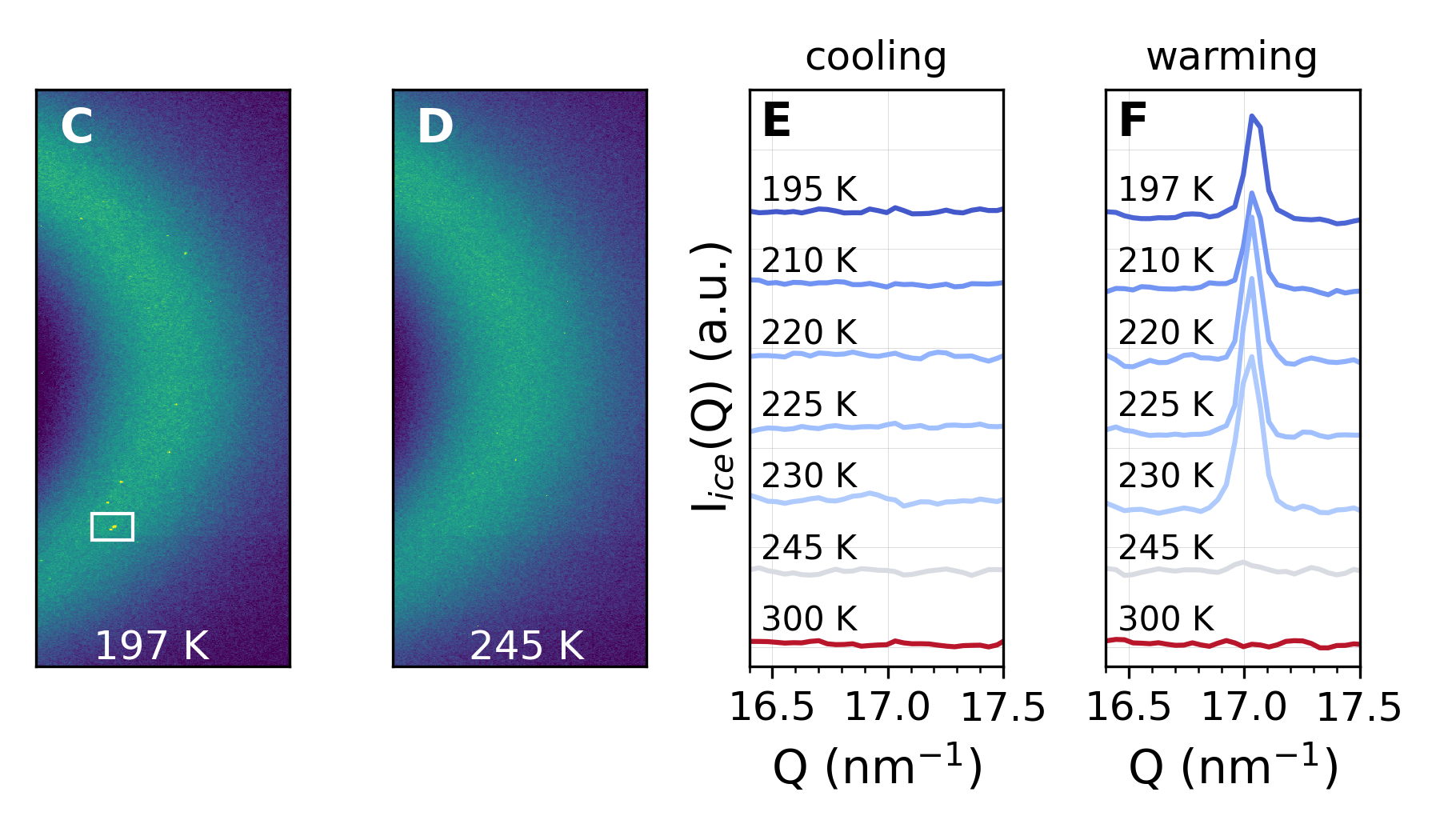}
 
  \caption{Wide-angle X-ray scattering intensities (WAXS) of a 200~mg/ml Lysozyme in 23 mol$\%$ glycerol-water solution as a function of temperature while cooling down from $T=300$~K to $T=195$~K (panel A) and warming back up to room temperature (panel B). The insets in both panels represent the data in contour plots to emphasize that the ice peaks are absent in the cool down and manifest in the warm up. The black arrows in panel B indicate the peaks matching some of the Bragg peaks of hexagonal ice. (C, D) representative 2D scattering patterns measured upon heating at $T=197$~K and $T=245$~K, where ice peaks appear in the former case. The white rectangle highlights the ice peak which profile along the radial direction is plotted in panels E and F upon cooling and warming, respectively. In panels E and F, an offset has been added to facilitate the comparison between temperatures. 
}\label{fig:waxs}
\end{figure}


Figure~\ref{fig:waxs} shows the variation of the WAXS scattering intensities of the 200~mg/ml Lysozyme glycerol-water solution upon cooling down to 195~K (panel A) and warming back to room temperature (panel B). The insets in both panels represent the data in contour plots zoomed in around the first peak in the $I(Q)$ where the ice signal is expected. The peaks in this $Q$-range are known to arise from scattering on interatomic length scales.

One can see that upon cooling, the $Q$ position of the first peak located at $Q_m^\text{WAXS} \approx 16.4$~nm$^{-1}$ at room temperature shifts to higher values and the peak shape narrows. 
The temperature trend observed for the glycerol-water mixture with the given glycerol concentration is opposite of what is know for pure water~\cite{chen_structures_2021}. Namely, while the $Q$ spacing between the first and the second peaks in water is known to increase with decreasing temperature, in Fig.~\ref{fig:waxs}, both peaks move together to higher $Q$ values.
We also note that even at the lowest temperatures (195~K), there are no detectable ice peaks, indicating that the system is not in its crystalline state even at such low temperatures thanks to the presence of glycerol. 

Upon heating, however, ice peaks start to be discernible in the WAXS intensities, as seen in Fig.~\ref{fig:waxs}B. Here, the black arrows indicate the $Q$ positions of several ice $I_h$ Bragg peaks. The 2D scattering patterns recorded by the detector at $T=197$~K and $245$~K are presented in panels C and D, respectively. One can see several sporadic ice Bragg peaks at $197$~K, which are already absent at $245$~K.  

For further analysis we focus on one of the ice Bragg peaks highlighted in Fig.~\ref{fig:waxs}C by a white rectangle and follow its evolution at different temperatures during cooling and heating. This peak corresponds to the ice $I_h$ $[002]$ diffraction peak centered at $\approx 17$ nm$^{-1}$ in the $I(Q)$ curves shown in panel B. We calculate the contribution from the ice peak by subtracting the diffuse part of the WAXS scattering. The resulting ice peak intensities $I_\text{ice}$ as a function of the momentum transfer $Q$ (i.e. radial direction of the 2D scattering patterns) are shown in Fig.~\ref{fig:waxs}F during heating. We note, that this ice peak is absent during the cooling down until 195~K (Fig.~\ref{fig:waxs}E).

One can see that the ice peak starts to develop already at $T\approx197$~K and disappears at $T\approx245$~K, which is slightly above the expected melting temperature for the 23 mol$\%$ glycerol-water mixture~\cite{murata_general_2013,popov_puzzling_2015}. From the width of the ice peak, we can estimate the ice crystallite size based on Scherrer's equation~\cite{langford_scherrer_1978,moreau_ice_2021}, $\delta = \frac{\lambda}{\beta \cos{\theta}}$, where $\delta$ is the apparent crystallite's size, $\lambda$ is the wavelength and $\beta$ is the breadth of the Bragg peak at scattering angle $\theta$. In this case, $\beta$ is calculated as the integral area beneath the peak divided by its maximum amplitude. Using this approach, we estimate that the ice crystallites starting from $197$~K are $\approx12-15$~nm in size. From the observed increase of the Bragg peak intensities during the warm up, the number density of these nanocrystallites also increases until $T\approx240-245$~K after which no ice peaks are visible anymore. The formation of ice nanocrystallites upon heating is attributed for pure glycerol-water mixture of similar concentrations to the consequence of the phase separation of the saturated glycerol-water domains and the excess water which eventually arranges into growing ice crystals~\cite{popov_puzzling_2015}.

We also note that under the present experimental conditions we are limited to the smallest size of the crystallites we can resolve in the order of $10$~nm. Below this, the peaks become very broad and comparable to the background~\cite{sellberg_ultrafast_2014}. Thus, we cannot exclude that the nanocrystals nucleate already during the cooling, although no ice Bragg peaks are observed (see Fig.~\ref{fig:waxs}A).

\begin{figure}[htbp]
  \centering
  \includegraphics[width=\columnwidth]{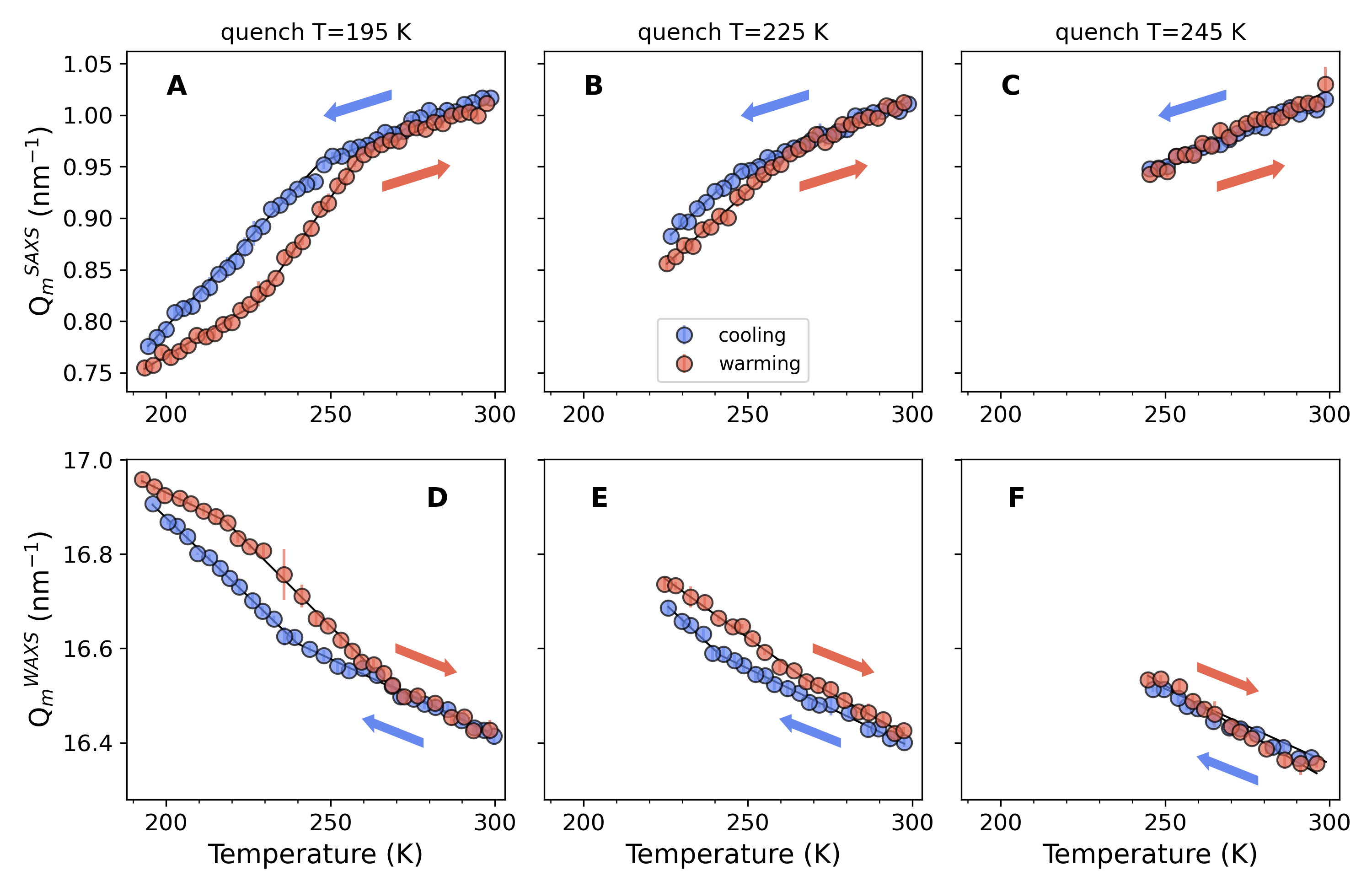}
  \caption{ Top row: temperature dependence of the $Q$-value of the SAXS $I(Q)$ peak position for a 200~mg/ml Lysozyme in glycerol-water solution during different temperature cycles: (A) deep quench with $T_\text{quench} = 195$~K, (B) medium with $T_\text{quench} = 225$~K and (C) shallow quench with $T_\text{quench} = 245$~K, respectively. Bottom row: simultaneously measured for the same sample the temperature dependence of $Q$-value of WAXS $I(Q)$ peak position during different temperature cycles: (D) deep, (E) medium and (F) shallow quench, respectively. The colors indicate measurement performed upon cooling (blue) or warming (red) as shown by the arrows.}\label{fig:tdep}
\end{figure}

 In Fig.~\ref{fig:tdep} (top row), we present the temperature dependence of the SAXS $I(Q)$ peak position $Q^\text{SAXS}_m$ as a function of temperature. 
 Interestingly, as seen in panel A the interference peak at $\approx1$~nm$^{-1}$ at room temperature in the SAXS $I(Q)$s corresponding to the protein-protein interactions shows different trends depending on whether the sample is being cooled (blue) or heated (red). Furthermore, the magnitude of the observed hysteresis depends on the final quench temperature. 
 Starting from room temperature, the peak gradually shifts towards lower $Q$ values until $T\approx245\pm1$~K. When cooling past this temperature, the slope increases significantly. Upon reheating the sample, however, the temperature dependence in Fig.~\ref{fig:tdep} exhibits a different path, corresponding to lower $Q_m^\text{SAXS}$ values for the same temperature. On the other hand, for the shallow quench in Fig.~\ref{fig:tdep}C, i.e. for $T_\text{quench}=245$~K, the hysteresis is reduced.

The bottom row in Fig.~\ref{fig:tdep} shows the temperature dependencies of the WAXS $I(Q)$ peak position $Q^\text{WAXS}_m$. Note that the WAXS data were measured simultaneously with the SAXS curves, i.e. during the same temperature cycles discussed above for the SAXS data. Similarly, when the sample is cooled down to $T=195$~K, a crossover in the temperature dependence of the peak position is observed at $T\approx 245$~K (Fig.~\ref{fig:tdep}D) with a significant hysteresis upon the full temperature sweep. The hysteresis becomes less apparent for the middle quench down to $T=225$~K (Fig.~\ref{fig:tdep}E) and disappears completely for the shallow quench to $T=245$~K (Fig.~\ref{fig:tdep}F).

We note that similar thermal hysteresis in the protein structure factor has been observed before~\cite{fitter_temperature_1999} and attributed to the formation of ice in hydration water for moderately hydrated protein powders. 

To elucidate the changes occurring in the system during the temperature variation, we study the protein-protein structure factor, $S(Q)$, extracted from the fits of the SAXS data (Fig.~\ref{fig:iqs}B). The obtained structure factor captures a pronounced shift of the low-$Q$ peak towards lower $Q$ values upon cooling observed in the experimental SAXS $I(Q)$ shown in Fig.~\ref{fig:iqs}B. The temperature evolution of the structure factor upon cooling the sample from room temperature down to $T=195$~K is shown in Fig.~\ref{fig:model}A. The TY-potential parameters used are fixed to $Z_1 = 21$, $Z_2 = 3.5$, and $K_2 = 3.2$ k$_\text{B}$T, and the interparticle attraction parameter $K_1$ is the fitting parameter. The $K_1$ variation in the whole temperature range (blue for cooling, red for warming up) and the resulting TY-potentials at each temperature are plotted in panels B and C, respectively. 

In agreement with the previous studies on Lysozyme solutions at ambient conditions~\cite{shukla_absence_2008,moller_effect_2012}, the effective attraction increases upon cooling  for the entire temperature range probed here. Furthermore, the temperature dependence of the $K_1$ parameter exhibits a crossover at $T\approx245$~K upon cooling, i.e. at a similar temperature where the crossovers in the experimental $I(Q)$s are observed (see Figs.~\ref{fig:tdep}A,D). Also here in red we present the extracted $K_1$ parameter from the fits of the SAXS data acquired during heating. Intuitively, the hysteresis shape resembles that obtained from the analysis of the peak in SAXS plotted in Fig.~\ref{fig:tdep}A.

To further elucidate the physical mechanism responsible for the observed changes of the protein-protein interaction at low-temperatures, we calculate the pair distribution function, $g(r)$, from the corresponding structure factor at different temperatures. The value of $g(r)$ describes the probability of finding another particle at a distance $r$ from the reference particle. Shown in Fig.~\ref{fig:model}D is the variation of $g(r)$ upon cooling from room temperature down to $195$~K. For all temperatures, when $r/\sigma<1$, $g(r) = 0$ consistent with the shape of the potential in panel C. The peak at $r/\sigma=1$ indicates the probability to find particles contact pairs owing to high protein concentration in the studied system.

Upon cooling the intensity of the first maximum at $r/\sigma \approx 1$ corresponding to the first coordination shell rises sharply resulting in a depletion in the range between $r/\sigma \approx 1.1$ and $r/\sigma \approx 1.2$. The latter also suggests that the interstitial protein molecules between the first and second coordination shell rearrange towards a more regular structure. Consistent with the observed increase in the attraction strength, the tendency for a higher probability density of first-neighbors is expected at lower temperatures. Furthermore, the second coordination maximum located at $r/\sigma \approx 1.5$ at room temperature gradually shifts towards larger distances with decreasing temperature. Eventually, at temperatures below $245$~K a pronounced feature in the $g(r)$ is developed at $r/\sigma \approx 2$, characteristic of finding an in-line configuration of three touching particles~\cite{riest_short-time_2018,liu_lysozyme_2011}. Overall, the observed behavior of the pair distribution function suggests an enhancement of the ordered arrangement of the protein molecules in the intermediate range upon cooling. 

\begin{figure}[htbp]
  \centering
\includegraphics[width=0.9\columnwidth]{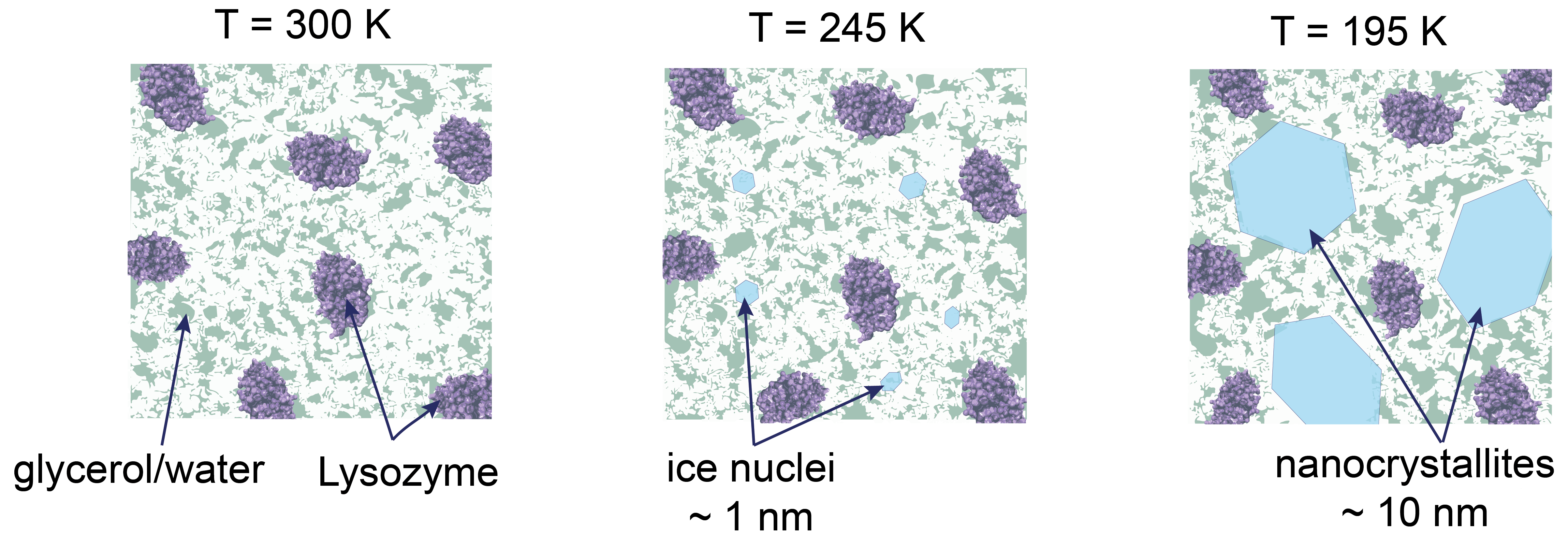}
  \caption{Schematic illustration of the changes occurring in the Lysozyme solution in glycerol-water mixture upon cooling below the melting point of the solvent (245 K), where ice crystallites start to nucleate, down to 195~K, where the nanocrystallites evolve further and grow upon reheating. }
  \label{fig:concept_figure}
\end{figure}

Assuming the observed crossover upon cooling below $\approx 245$~K occurs due to ice nanocrystallite formation (Fig.~\ref{fig:concept_figure}), it can be deduced that the glycerol concentration in the remaining solvent slightly changes due to the expelled glycerol from the ice\cite{hayashi_ice_2005,hayashi_relaxation_2005,popov_puzzling_2015}. As a result, the protein-protein interactions below this temperature occur in an effectively different environment at colder temperatures. This interpretation is consistent with the observed increase of the slope of the attraction strength $K_1$ after the crossover temperature, which can be related to the decrease of the dielectric permittivity of the medium\cite{javid_proteinprotein_2007}.

Furthermore, during the formation of the ice crystallites, the structure of the solvation layer, i.e. the layer around the protein, is expected to change as well. Previous studies suggest that in aqueous solutions, as the glycerol concentration increases above 50 vol$\%$, glycerol molecules are more included into the protein solvation shell~\cite{hirai_direct_2018} which otherwise is highly disfavoured from the vicinity of the protein~\cite{sinibaldi_preferential_2007,gekko_mechanism_1981}.
In turn, the composition of the surface layer can affect the protein-protein interactions~\cite{shulgin_various_2008} which in our case manifests in the kink in the observed temperature dependence of the attraction strength $K_1$.
Consequently, the restructuring is reflected in the changes in the pair distribution function and the position of the correlation peak in SAXS.

\section{CONCLUSIONS}

In conclusion, we present a simultaneous SAXS/WAXS study of the Lysozyme solution in glycerol-water mixture in a broad temperature range from room temperature to $\approx195$~K. We follow the temperature evolution of the protein-protein peak in SAXS as well as the interatomic structure peak in WAXS upon cooling down and heating. The hysteresis observed in both SAXS and WAXS upon the full temperature cycle is attributed to the formation of nanocrystallites with the size of $\approx10$~nm, which are evident by the Bragg peaks in the WAXS during heating. Furthermore, the observed crossover at $245$~K upon cooling is found both in SAXS and WAXS, which coincides with the melting temperature of the solution. We attribute this effect to the ice nuclei formation occurring already upon cooling down below the melting point of the solutions.

 Since we do not observe any significant temperature dependent changes in the protein radius of gyration $R_g$, we tentatively conclude that this transition does not reflect cold denaturation. Instead, we attribute this transition to changes in the protein-protein interaction potential stemming from the influence of the solvent due to the nanocrystals, which is modeled by the two-Yukawa potential. The model indicates that this effect is reflected in the protein-protein structure factor peak and results in increased protein-protein attraction. 
From the observed variation of the pair distribution function, we infer that upon cooling the interstitial range in the probability density between first and second protein-protein coordination shell is depleted due to the increased attraction term between the Lysozyme molecules.

Our results shed light on the influence of nanocrystalites on protein-protein interactions and functionally mechanisms of cryoprotectants at low temperatures. These insights can advance our understanding of protein stability in supercooled environments, with important implications for biotechnical cryostorage applications and for the development of new technologies and materials to improve our ability to survive and thrive in cold and icy environments.

\section{ACKNOWLEDGEMENTS}
We acknowledge financial support by the Swedish National Research Council (Vetenskapsr\text{\aa}det) under Grant No. 2019-05542 and within the R\"{o}ntgen-\text{\AA}ngstr\"{o}m Cluster Grant No. 2019-06075. This research is supported by Center of Molecular Water Science (CMWS) of DESY in an Early Science Project, the MaxWater initiative of the Max-Planck-Gesellschaft and the Wenner-Gren Foundations. C.G. and S.T. acknowledge the support from Bundesministerium f\"ur Bildung und Forschung (BMBF) with grants No. 05K19PS1, 05K20PSA and 05K22PS1. Parts of this research were carried out at the light source PETRA III at DESY, a member of the Helmholtz Association (HGF). We acknowledge the European Synchrotron Radiation Facility (ESRF) for provision of synchrotron radiation facilities, and we thank Theyencheri Narayanan and Thomas Zinn for assistance with preliminary measurements at the beamline ID02.

\bibliography{My_Library.bib}

\end{document}